\documentclass[9pt,twocolumn,twoside]{osajnl}

\journal{ao} 

\setboolean{shortarticle}{false}

\usepackage{lineno}

\usepackage{amsmath}
\usepackage{tikz}
\usetikzlibrary{arrows, arrows.meta}
\usepackage{graphicx,nicefrac}
\usepackage{multirow}
\usepackage{booktabs}
\usepackage{stfloats}
\usepackage{svg}
\usepackage[short]{optidef}
\usepackage{bm}
\usepackage{algorithm}
\usepackage{algpseudocode}

\newcommand\norm[1]{\left\lVertX1\right\rVert}
\newcommand\norml[1]{\left\lVertX1\right\rVert_{1}}
\newcommand\normll[1]{\left\lVertX1\right\rVert^{2}_{2}}

\title{JR2net: A Joint Non-Linear Representation and Recovery Network for Compressive Spectral Imaging}

\author[$\!$]{Brayan Monroy}
\author[$\!$]{Jorge Bacca}
\author[*]{Henry Arguello}

\affil[$\!$]{Department of Systems Engineering, Universidad Industrial de Santander, Bucaramanga, Colombia}

\affil[*]{Corresponding author: henarfu@uis.edu.co}

\begin{abstract}
Deep learning models are state-of-the-art in compressive spectral imaging (CSI) recovery. These methods use a deep neural network (DNN) as an image generator to learn non-linear mapping from compressed measurements to the spectral image. For instance, the deep spectral prior approach uses a convolutional autoencoder network (CAE) in the optimization algorithm to recover the spectral image by using a non-linear representation. However, the CAE training is detached from the recovery problem, which does not guarantee optimal representation of the spectral images for the CSI problem. This work proposes a joint non-linear representation and recovery network (JR2net), linking the representation and recovery task into a single optimization problem.  JR2net consists of an optimization-inspired network following an ADMM formulation that learns a non-linear low-dimensional representation and simultaneously performs the spectral image recovery, trained via the end-to-end approach. Experimental results show the superiority of the proposed method with improvements up to 2.57 dB in PSNR and performance around 2000 times faster than state-of-the-art methods.
\end{abstract}

\setboolean{displaycopyright}{true}

\begin{document}

\maketitle

\section{Introduction}

\begin{figure*}[!t]
    \centering
    \includegraphics[width=\linewidth]{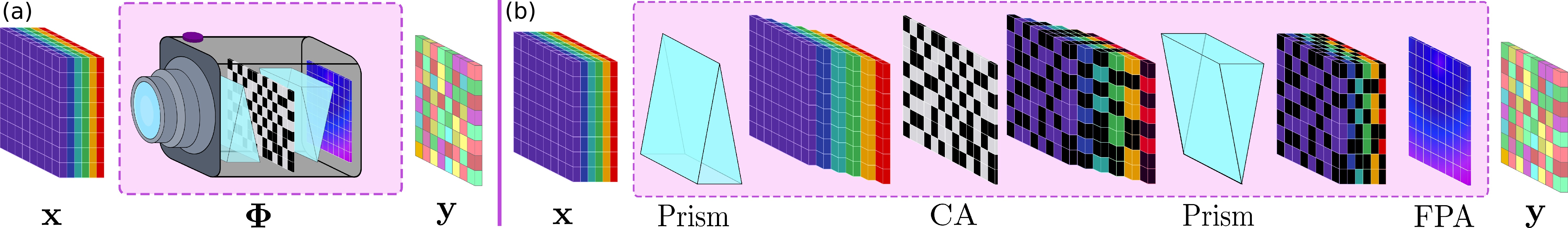}
    \caption{ (a) Dual Disperser DD-CASSI system. (b) The input spectral
image $\textbf{x}$ is sheared along the spectrum. Subsequently, the information is coded following the transmission function of the
coded aperture (CA), creating a coded signal scene. The coded signal is unsheared, and finally, a focal plane array (FPA) is used to
obtain the 2D compressed projections.} 
    \label{fig:ddcassi}
\end{figure*}

\textcolor{black}{Spectral imaging senses spatial information along several wavelengths of the electromagnetic spectrum, which combines imaging and spectroscopy. Imaging consists of obtaining the spatial information of the scene by measuring the intensity of light emitted by objects. On the other hand, spectroscopy consists of sensing the spectral information of the object by measuring the response on the electromagnetic spectrum along multiple wavelengths \cite{garini2006spectral}. \textcolor{black}{Combining} of these sub-fields enables spectral imaging systems to acquire spatial-spectral information of objects, which can be represented as a 3D data cube where two dimensions stand for the spatial information and the third for the spectral information \cite{bacca2019sparse}. Due to the importance of spectral information presented in objects, spectral images have been used in precision agriculture \cite{agriculture}, medical imaging \cite{medical}, remote sensing \cite{remote}, and microscopy \cite{garini2006spectral}. }

Several scanning methods have been explored for their potential to acquire spectral images. For instance, techniques such as whiskbroom, pushroom, and filtering changing sense spectral pixels, spectral lines, or spectral bands, respectively. However, these systems sequentially sense a subset of the data cube and thus require large sensing time to obtain a complete data cube~\cite{wagadarikar2008single,gao2014single,liang2018single}. Hence, snapshot techniques have been proposed based on compressive sampling (CS) theory \cite{csi}. For instance, snapshot spectral imaging systems such as coded aperture snapshot spectral imager (CASSI) based on single-disperser (SD-CASSI)~\cite{wagadarikar2008single} or dual-disperser (DD-CASSI)~\cite{ddcassi}, employs coded apertures and dispersive elements to modulate the optical path from a scene. A detector captures a two-dimensional compressed projection of the three-dimensional data cube, reducing the amount of captured data~\cite{wagadarikar2008single}. This snapshot system describes an under-determined system where a recovery step is needed to recover an estimation of the 3D data cube from the compressed projection~\cite{bacca2021deep}.

Compressive spectral optimization algorithms assume some prior information about the natural signal to solve the under-determined recovery problem and reduce the number of parameters to be recovered. For instance, due to the redundant spatial and spectral information presented in the scene,  algorithms often assume a sparse representation on some orthonormal basis, such as Wavelet, discrete cosine transform (DCT), and Kronecker-Wavelet~\cite{csi,fu2016exploiting, wang2016adaptive, duarte2011kronecker}. On the other hand, deep learning-based methods aim to learn a spectral prior as a non-linear representation of the spectral images, outperforming sparse representation techniques. For instance, Autoencoder~\cite{kaist, monroy2021deep} uses a convolutional autoencoder (CAE) to learn a non-linear representation of the spectral data in an off-line training, which is later included as spectral prior in the optimization algorithm. However, the choice of the linear representation or the learning of the non-linear representation is not designed for the recovery problem where it is finally used; therefore, it does not guarantee an optimal recovery of spectral data in the CSI problem.

Consequently, this paper presents JR2net, a joint non-linear representation and recovery network for compressive spectral imaging. JR2net is inspired by an ADMM formulation that links the representation and recovery problem into a single optimization problem. Each ADMM iteration is unrolled into a network stage that involves a convolution decoder to learn the non-linear low-dimensional (NLD) representation shared along the stages. Furthermore, this work proposes a learned gradient network to learn the derivative of the convolutional decoder, and we present a loss function inspired by CAEs incorporated into the end-to-end training procedure. Experimental results show the superiority of the proposed method with improvements up to 2.57 dB in PSNR and performance around 2000 times faster than state-of-the-art methods.
	    {
	           The main contributions of the proposed  framework are summarized in the following items.}
        \begin{itemize}
        	    \item {JR2net: A joint non-linear representation and recovery network  for compressive spectral imaging.}
	           \item {A learned gradient network $\bm{G}(\cdot)$: We proposed to learn a neural network to replace the gradient of the convolutional decoder in an unrolled formulation.}
	           \item {An autoencoder-inspired loss function:  Based on the fact that the proposed gradient network has the spatial dimensions of an encoder, we incorporated the loss function used in the training of CAEs in the end-to-end training procedure.}
	     \end{itemize}

\section{Related Work}
\label{sec:related_work}

\subsection{Model-based Optimization}
Traditional CS recovery algorithms are considered to be hand-crafted as they use expert knowledge of the signal as prior information. These methods are based on optimization techniques that consist of a data fidelity term and incorporate the hand-crafted prior as a $l_0$ or $l_1$ regularization term \cite{figueiredo2007gradient}. To name, \cite{tv,twist,wagadarikar2008single} assume low total variation (TV), \cite{csi,correa2015snapshot} explore spatial sparsity on a given basis such as Wavelet and spectral sparsity DCT domain \cite{csi,fu2016exploiting, wang2016adaptive} or \cite{bacca2019noniterative,gelvez2017joint} employ low-rank structures based on the linear mixture model. Examples of algorithms that use sparsity priors include the GPSR \cite{figueiredo2007gradient}, ADMM  \cite{boyd2011distributed}, CSALSA \cite{afonso2010augmented}, ISTA \cite{beck2009fast}, and AMP \cite{donoho2009message} among others. In these algorithms, an auxiliary variable is incorporated in the optimization problem to alternately minimize the data fidelity term and the $l_0$ or $l_1$ regularization term, where the latter is solved by a \textit{projection operator} \cite{figueiredo2007gradient,donoho2009message} or a \textit{proximal operator} \cite{boyd2011distributed,afonso2010augmented,beck2009fast}. However, these hand-crafted methods require expert knowledge of the scene to select which prior is more appropriate for this spectral scene. Consequently, they do not represent the wide variety and non-linearity of spectral image representations~\cite{dip_bacca}. 

\subsection{Model-based Optimization with Deep Priors}
These methods are based on iterative techniques that replace the hand-crafted prior with a deep neural network (DNN) used to learn a deep prior of the spectral images. Non-data-driven approaches \cite{deepPrior, dip_bacca} proposed untrained DNN as a deep generative model (DGM) where the input of the network (denoted as latent space) passes through convolution operators to generate the image recovery. The weights of untrained DNN \textcolor{black}{aim to minimize the Euclidean distance between} the forward sensing operator of the DNN output and the compressed measurement. Plug-and-Play with convolutional neural networks (PnP-CNNs) replace the proximal operator with a pre-trained CNN \cite{yuan2020plug,chan2016plug,rick2017one}. Finally, Autoencoder \cite{kaist} pre-trained the DGM to learn a non-linear representation using a convolutional autoencoder which is then incorporated into the optimization problem. These methods replace the expert knowledge of the target with extensive use of data. Additionally, these iterative methods need a training step for the non-linear representation learning, which does not contemplate the optimization problem or the particular sensing system.

\subsection{Deep-based Optimization}
These methods propose a DNN, which performs the inverse mapping from the compressed measurements to the spectral image. For instance, \cite{gedalin2019deepcubenet} proposed a U-net-based network as non-linear mapping replacing the 2D for 3D convolutions, \cite{mousavi2015deep} learns a stacked auto-encoder, \cite{mousavi2017learning} applied convolution layers, and convolutional, residual, and fully connected layers are also used in \cite{dave2018solving,palangi2016distributed,yao2019dr2,kulkarni2016reconnet}. Authors in {\cite{meng2020end} uses the attention mechanism in order to jointly capture self-attention across different dimensions in spectral images, \cite{wang2021simple} includes a residual learning strategy with nested structures and spatial-invariant property.}  However, changes in the sensing system can drastically affect the performance of these methods. Recently, a novel approach suggested unrolling the traditional optimization method and learning the proximal operators resulting in a deep interpretable decoder \cite{kaist2}, with some authors, such as \cite{dnu, kaist2, sogabe2020admm} replacing the soft-thresholding operator for a spatial-spectral network. These methods allow learning the optimization parameters within the weights network using an end-to-end optimization \cite{bacca2021deep,sitzmann2018end,arguello2022deep}. Although learning the proximal operator increases the robustness of the method against changes in the sensing system, it depends on the particular optimization problem, limiting its use in other representation problems, unlike non-linear representation learning methods~\cite{kaist}.

\begin{figure*}[!t]
    \centering
    \includegraphics[width=\textwidth]{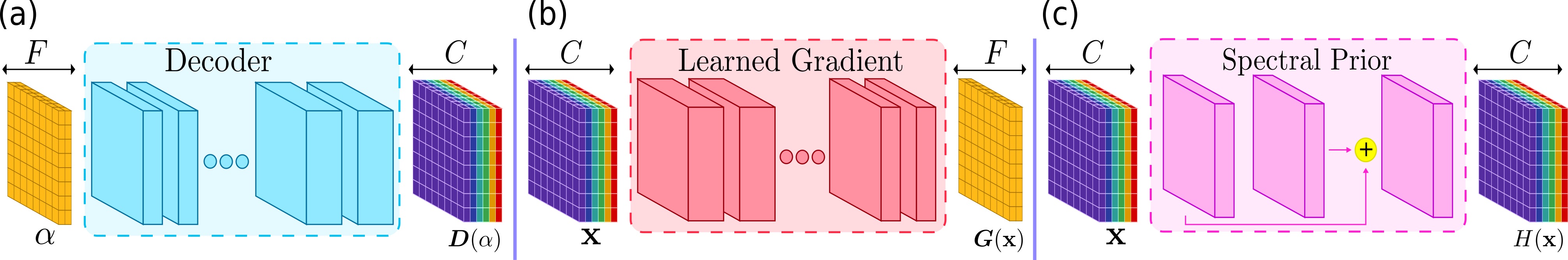}
    \caption{ (a) Decoder network and (b) learned gradient network composed of $d$-layered convolutional blocks. (c) The spectral prior network consists of three convolutional blocks with a skip connection. These three interpretable blocks were incorporated into the recovery algorithm.}
    \label{fig:networks} \vspace{-1em}
\end{figure*} 

\section{Compressive Spectral Imaging}  

Compressive spectral imaging (CSI) encodes and projects the spatial-spectral information of the scene through optical elements, as shown in Fig.~\ref{fig:ddcassi}. For instance, the dual-disperser (DD-CASSI) architecture comprises two main optical elements: the coded aperture and the prism. Physically, a spectral scene $x(i,j,k)$ is first collected by the objective lens, which is sheared along the spectrum by a prism, and then spatially en\textbf{}coded by a coded aperture~$T(i,j)$. Subsequently, the encoded sheared scene is unsheared through a second prism, and finally, a focal plane array (FPA) is used to obtain the 2D compressed projections $y(i,j)$. The resulting 2D compressed projections can be mathematically described as
\begin{equation}    
    y_{(i,j)} =\sum_{k=1}^{C} T_{(i+C-k,j)}x_{(i,j,k)}.
    \label{eqn:sd_integral}
\end{equation}
Eq.~(\ref{eqn:sd_integral}), can be re-written in matrix-vector form as

\begin{equation}    
    \textbf{y} =\bm{\Phi} \textbf{x} + \boldsymbol{\eta},
    \label{eqn:integral}
\end{equation}
where $\textbf{y} \in \mathbb{R}^{HW}$ and $\textbf{x} \in \mathbb{R}^{HWC}$ are the vectorized representation of the 2D compressed projections and the spectral scene, respectively, with $H, W$ as the spatial size, and $C$ as the number of spectral bands, $ \bm{\Phi} \in \mathbb{R}^{HW \times HWC}$ is the sensing matrix that describes the DD-CASSI image acquisition model, and $ \bm{\eta} \in \mathbb{R}^{HW}$ is the noise corruption inherent in the acquisition model. This equation describes a highly under-determined system since $HW << HWC$.

\section{Proposed Method}

The proposed jointly representation and recovery method consist of an unrolled network inspired by an ADMM formulation. The unrolled network comprises three sub-blocks of interpretable layers, the decoder, the learned gradient network, and the spectral prior, as shown in Fig.~\ref{fig:networks}. The interpretable blocks are integrated into the optimization problem, where each specializes in solving a specific problem. The decoder $\bm{D}(\cdot)$ expands the NLD representation to the original spectral image. The learned gradient network $\bm{G}(\cdot)$ learns the derivatives of the $\bm{D}(\cdot)$ and simultaneously performs the non-linear transformation from the spectral images domain to the NLD representation domain. Finally, the spectral prior network $\bm{H}(\cdot)$ learns a data-driven prior to further improve the spectral fidelity by substituting the proximal operator, which is explained in detail below.

\subsection{Representation Learning}

The representation learning consists in learn a non-linear representation $\bm{\alpha} \in \mathbb{R}^{HWF}$ from a given spectral image $\textbf{f} \in \mathbb{R}^{HWC}$, via the learning of a \textit{decoder} $\bm{D}(\cdot)$ which are composed of a set of convolutional layers and activations operators, as shown in Fig. \ref{fig:networks}(a). The decoder network expands the NLD representation to produce the original input data $\textbf{x} \approx \bm{D}(\bm{\alpha})$. Using the decoder network, the compressive image formation can be re-written as
\begin{equation}
      \textbf{y} = \bm{\Phi} \textbf{x} \approx \bm{\Phi D}(\bm{\alpha}),
\end{equation}
    Notice that when $HWF < HWC$ $, \bm{\alpha}$ is an NLD representation of the spectral image. {In this work, we use the decoder network presented in~\cite{monroy2021deep}, which will be later trained to guarantee an NLD representation.}

\begin{algorithm}[!t]
\caption{ADMM Solution for the optimization problem in Eq.~(\ref{eqn:reform})}
\label{alg:admm_unrolled}
\hspace*{\algorithmicindent} \textbf{Input: } $\bm{\Phi} , \textbf{y}$
\label{euclid}\begin{algorithmic}[1]
\State $\bm{\alpha}^{(0)}  =  \bm{G(\bm{\Phi}^{\top}\textbf{y})}$ \Comment{LD representation initialization}
\State $\textbf{u}^{(0)} = 0$ \Comment{auxiliary variable initialization}
\While{not convergence}
\State $\nabla_{\bm{\alpha}}\mathcal{L}= \bm{G}( \bm{\Phi^{\top} \Phi D}(\bm \alpha) - \bm{\Phi}^{\top}\textbf{y} ) + \rho \bm{G}( \bm{D}(\bm{\alpha}) - \textbf{h} + \textbf{u})$  
\State  $ \bm{\alpha}^{(k+1)} = \bm{\alpha}^{(k)} - \mu \nabla_{\bm{\alpha}}\mathcal{L} $
\State $\textbf{h}^{(k+1)} = \bm{H}^{(k+1)}(\bm{D}(\bm{\alpha}^{(k+1)}) + \textbf{u})$ 
\State $\textbf{u}^{(k+1)} = \textbf{u} + \bm{D}(\bm{\alpha}^{(k+1)}) - \textbf{h}^{(k+1)}$
\EndWhile
\State $\hat{\textbf{x}} = \bm{D}(\bm{\alpha}^{(k+1)})$ \Comment{Spectral image recovery}
\State \textbf{return} $\hat{\textbf{x}}$
\end{algorithmic}
\end{algorithm}

\begin{figure*}[!t]
    \centering
    \includegraphics[width=\linewidth]{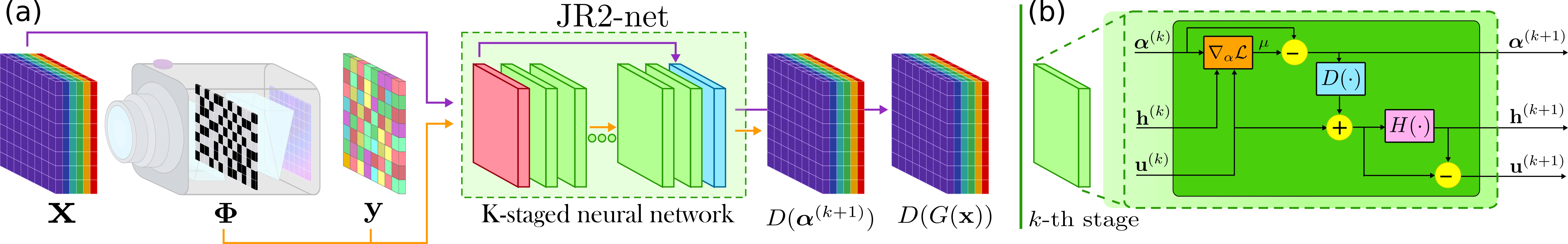}
    \caption{ (a) The proposed joint representation and recovery network architecture (JR2net). (b)~$k$-th stage into the JR2net, each stage represents an iteration of the Algorithm~\ref{alg:admm_unrolled}. For the first stage, $\bm{\alpha}^{(0)}$ is initialized as $\bm{G}(\bm{\Phi}^{\top}\textbf{y})$. }
    \label{fig:jr2net} 
\end{figure*}

\subsection{Recovery Method}

In order to recover the spectral image from the compressive measurements taking into account the NLD representation,  we propose an unrolling method inspired by the following optimization problem
\begin{equation}
  \underset{\bm{\alpha}}{\text{minimize} }\quad \Vert \textbf{y} - \Phi \bm{D}(\bm{\alpha})\Vert^2_2 + \tau R(\bm{D}(\bm{\alpha})),
\label{eqn:opti}
\end{equation} where the first term is the data fidelity term, $R$ is a prior term, and $\tau$ is the regularization parameter that controls the relative importance of the regularization term. The optimization problem in Eq.~(\ref{eqn:opti}) can be written as
\begin{equation}
\begin{aligned}
  &\underset{\bm{\alpha} , \textbf{h}}{\text{minimize} }\quad \bm{f}(\bm{\alpha}) + \bm{g}(\textbf{h}) , \\ 
  &\text{subject to} \quad \bm{D}(\bm{\alpha}) = \textbf{h},
 \end{aligned} \label{eqn:reform}
\end{equation}
where $\bm{f}(\bm{\alpha}) = \Vert \textbf{y} - \Phi \bm{D}(\bm{\alpha})\Vert^2_2  $ and $\bm{g}(\textbf{h})= \tau R(\bm{D}(\bm{\alpha})) $. Then, given the augmented Lagrangian in Eq.~(\ref{eqn:reform}) as 
\begin{equation}
    \mathcal{L}(\bm{\alpha}, \textbf{h}) = \bm{f}(\bm{\alpha}) + \bm{g}(\textbf{h}) + \frac{\rho}{2} \Vert  \bm{D}(\bm{\alpha}) - \textbf{h}  + \textbf{u} \Vert_2^2, \label{eqn:lagrangian}
\end{equation} where $\textbf{u}$ is an auxiliary variable, this problem can be solved using the alternating direction method of multipliers (ADMM), as shown in Algorithm~\ref{alg:admm_unrolled}. ADMM scheme consists of iteratively minimizing Eq.~(\ref{eqn:lagrangian}) via the alternating solution of $\bm{\alpha}, \textbf{h}$ subproblems. The key points of this solution are summarized below. First, the $\bm{\alpha}$-subproblem can be expressed as
\begin{equation}
      \bm{\alpha}^{(k+1)} = \underset{\bm{\alpha}}{\text{arg min}} \quad \bm{f}(\bm{\alpha}) + \frac{\rho}{2} \Vert  \bm{D}(\bm{\alpha}) - \textbf{h}  + \textbf{u} \Vert_2^2 \label{eqn:min_alpha},
\end{equation} due to fact that $\bm{D}$ is a non-linear operator, it is not feasible to obtain a closed-form solution for Eq.~(\ref{eqn:min_alpha}). Instead, Eq.~(\ref{eqn:min_alpha}) is solved using a gradient descent algorithm, following the gradients $\nabla_{\bm{\alpha}} \mathcal{L}$ of the $\bm{\alpha}$-subproblem as
\begin{equation} 
\begin{aligned}
        \nabla_{\bm{\alpha}}\mathcal{L} = &\nabla_{\alpha}\bm{D}(\bm{\alpha})\cdot(  \bm{\Phi^{\top} \Phi D}(\bm \alpha) - \bm{\Phi}^{\top}\textbf{y} ) \\ &+ \rho  \nabla_{\alpha}\bm{D}(\bm{\alpha})\cdot( \bm{D}(\bm{\alpha}) - \textbf{h} + \textbf{u} ), \label{eqn:gradlagrangian}
\end{aligned}
\end{equation}
where $\bm{\Phi}^{\top}\textbf{y}$ represents the initialization. 

It is worth noting that the above expression involves computing the derivative of the decoder network with respect to the input $\nabla_{\alpha}\bm{D}(\boldsymbol{\alpha})$, which results in an expensive matrix calculation of $\mathbb{R}^{HWF\times HWC}$ dimensions. {Therefore, we propose learning $\nabla_{\bm{\alpha}}\bm{D}(\bm{\alpha})$ with the learned gradient network $\bm{G}(\cdot)$, which follows the architecture of a convolutional encoder as shown in Fig.~\ref{fig:networks}(b). In that manner, $\bm{G}(\cdot)$ aims to learn the derivative of the decoder network. More details on CAE can be found in \cite{monroy2021deep}. Including the learned gradient network, } Eq.~(\ref{eqn:gradlagrangian}) can be re-written as


\begin{equation}
       \nabla_{\bm{\alpha}}\mathcal{L} \approx   \bm{G}( \bm{\Phi^{\top} \Phi D}(\bm \alpha) - \bm{\Phi}^{\top}\textbf{y} ) +  \rho \bm{G}( \bm{D}(\bm{\alpha}) - \textbf{h} + \textbf{u} ), \label{eqn:alpha2}
\end{equation}
then, following the gradient descent update rule, the solution to Eq.~(\ref{eqn:min_alpha}) can be expressed as 

\begin{equation}
    \bm{\alpha}^{(k+1)} = \bm{\alpha}^{(k)} - \mu  \nabla_{\bm{\alpha}}\mathcal{L},
\end{equation} where $\mu$ is the step size in the gradient descent. Second, the $\textbf{h}$-subproblem can be expressed as \begin{equation}
\begin{aligned}
    \textbf{h}^{(k+1)} &= \underset{\bm{\alpha}}{\text{arg min}} \quad \bm{g}(\textbf{h}^{(k)}) + \frac{\rho}{2}  \Vert  \bm{D}(\bm{\alpha}^{(k+1)}) - \textbf{h}  + \textbf{u} \Vert_2^2, \\
    &= \underset{\bm{\alpha}}{\text{arg min}} \quad \tau R(\textbf{h}^{(k)}) + \frac{\rho}{2}  \Vert  \bm{D}(\bm{\alpha}^{(k+1)}) - \textbf{h}  + \textbf{u} \Vert_2^2
    \label{eqn:hproblem} ,
\end{aligned}
\end{equation} it can see that solution to Eq.~(\ref{eqn:hproblem}) is a proximal operator of the spectral image prior $R$ with penalty $\tau$ \cite{boyd2004convex}. When the spectral image prior uses the $l_1$ sparsity, the proximal operator implies a soft-thresholding on $\bm{D}(\bm{\alpha}^{(k+1)} + \textbf{u})$. In this work, instead of explicitly learning a spectral image prior $R$ and solving the operator with the constraint of the spectral image prior, we adopt the approach used in \cite{dnu, kaist2} and learn a solver $\bm{H}(\cdot)$ for the proximal operator with a spectral prior network, expressed as follows
\begin{equation}
    \textbf{h}^{(k+1)} = \bm{H}(\bm{D}(\bm{\alpha}^{(k+1)}) + \textbf{u}).
\end{equation} The spectral prior network  $\bm{H}(\cdot)$ is a CNN composed of three convolutional blocks with a skip connection, as shown in Fig.~\ref{fig:networks}(c). Finally, the auxiliary variable update
\begin{equation}
      \textbf{u}^{(k+1)} = \textbf{u}^{(k)} +  \bm{D}(\bm{\alpha}^{(k+1)}) - \textbf{h}^{(k)}.
\end{equation}

\subsection{Joint Non-Linear Representation-Recovery Network}

The ADMM scheme consists of iteratively applying the steps summarized in Algorithm~\ref{alg:admm_unrolled} until a stop criterion is satisfied. The unrolling technique consists of unrolling $K$-iterations of the optimization algorithm into $K$ stages of a DNN, in which each stage corresponds to one iteration in the optimization algorithm. This technique allows us to jointly learn both the network and optimization parameters, stabilizing training and improving the speed and accuracy of spectral images recovered~\cite{sogabe2020admm, liu2018proximal}.

To learn the NLD representations and the recovery problem simultaneously, the Algorithm~\ref{alg:admm_unrolled} is unrolled into a $K$-staged network as shown in Fig.~\ref{fig:jr2net}. In this architecture, each $k$-stage of the unrolled network are composed of the $\bm{\alpha}^{(k)}, \textbf{h}^{(k)}, \textbf{u}^{(k)}$ update steps. In order to learn a unique deep spectral representation, the sub-networks $\bm{G}(\cdot)$ and  $\bm{D}(\cdot)$ are shared along the stages, which significantly reduces the parameters of the unrolled network. All the parameters in each stage are individually defined stage by stage, and $\mu^{(k)}$, $\rho^{(k)}$ and $\bm{H}^{(k)}(\cdot)$ represent $\mu$, $\rho$ and $\bm{H}(\cdot)$ at $k$-th stage.   Once the network is built, {we perform the end-to-end training with the training set of $N$ spectral patches and synthetic sensing matrices   $\{ \textbf{x}^{(i)}, \bm{\Phi}^{(i)} \}_{i=1}^N$}, to learn the optimization parameters $ \Omega = \{ \mu^{(k)}, \rho^{(k)} \}_{i=1}^{K}$, the NLD representations and the network parameters for the recovery problem. Therefore, the complete objective for finding the optimal set of model parameters $\bm{Q}^* = \{\bm{D}^*, \bm{G}^*, \bm{H}^{*(k)},\Omega^*\}$ is given by
\begin{equation}
    \bm{Q}^* = \underset{ \bm{Q} }{\text{arg min}} \quad   \mathbb{E}\Big[ \Vert \hat{\textbf{x}}^{(i)} - \textbf{x}^{(i)}  \Vert_2^2 \Big], \label{eqn:loss_train}
\end{equation} where  $\hat{\textbf{x}}=\bm{U}(\textbf{y}, \bm{\Phi}; \bm{Q})$ is the recovered spectral image from the unrolled network $\bm{U}(\cdot)$ given the input sensor measurements $\textbf{y}$, sensing matrix $\bm{\Phi}$, and the model parameters $\bm{Q}$.

\subsection{Autoencoder Loss Function}

CAEs consist of an \textit{encoder} $\bm{E}(\cdot)$ and a \textit{decoder} $\bm{D}(\cdot)$, where each operator is composed of a set of convolutional layers and activation operators \cite{masci2011stacked}. The encoder network extracts the features and transforms the input data into a non-linear representation $\bm{\alpha} = \bm{E}(\textbf{x})$, then the decoder network expands the NLD representation to produce the original input data $\textbf{x} \approx \bm{D}(\bm{\alpha})$. This type of network has been successfully applied in tasks such as image classification \cite{zhang2016augmenting}, image clustering~\cite{ghasedi2017deep}, and denoising \cite{gondara2016medical}. The forward model of the CAE can be mathematically expressed as
\begin{equation}
    \bm{A}(\textbf{x}) = \bm{D}({\bm{E}(\textbf{x}))}  \approx \textbf{x}. \label{eqn:autoencoder}
\end{equation}
	{Note that both the convolutional encoder and the learned gradient network $\bm{G}(\cdot): \mathbb{R}^{MNC} \rightarrow \mathbb{R}^{MNF} $, perform a non-linear transformation from the spectral images domain to the NLD domain $\bm{\alpha}$. Therefore, inspired by the CAEs architectures, we introduce the autoencoder loss function $ \bm{L}_{ae}$, which aims to train the $\bm{G}(\cdot)$ network as an encoder network by measuring the error between the original and the autoencoder representation spectral images as
	\begin{equation} 
        \bm{L}_{ae} =  \Vert \textbf{x}^{(i)} - \bm{D}(\bm{G}(\textbf{x}^{(i)}) )\Vert_2^2.
    \end{equation}
	}
We would like to highlight that instead of pre-training a convolutional autoencoder into a spectral image dataset and then training a second network for the image recovery, the proposed model jointly learns the NLD representations of spectral images and recovers them from these representations; in this sense, the $\bm{G}(\cdot)$ aims to learn the derivatives of the decoder $\nabla_{\alpha}\bm{D}(\cdot)$ and the non-linear mapping from the NLD representation at the same time. This can be done by including the autoencoder representation loss $\bm{L}_{ae}$ into the training procedure.

\subsection{Computational Complexity}

With regard to the computational complexity of the Algorithm~\ref{alg:admm_unrolled}. The most costly steps are the forward pass of $\bm{G}(\cdot)$ in Line 4, which are the order of complexity $\mathcal{O}(HWF_{max})$, where $F_{max}$ is the convolutional layer with a maximum number of filters. The matrix multiplication by $\bm{\Phi}$ and $\bm{\Phi}^{\top}$ can be performed element-wise due to definition in Eq.~(\ref{eqn:sd_integral}), and thus has an order of complexity $\mathcal{O}(HWC)$. Therefore, the overall order of complexity per iteration is given by $\mathcal{O}(HWF_{max}) + \mathcal{O}(HWC)$. Finally, Algorithm~\ref{alg:admm_unrolled} has computational complexity $\mathcal{O}(N_kHWF_{max}) + \mathcal{O}(N_kHWC)$ where $N_k$ is the number of stages in the unrolled network.

\begin{figure}[!t]
    \centering
    \includegraphics[width=\linewidth]{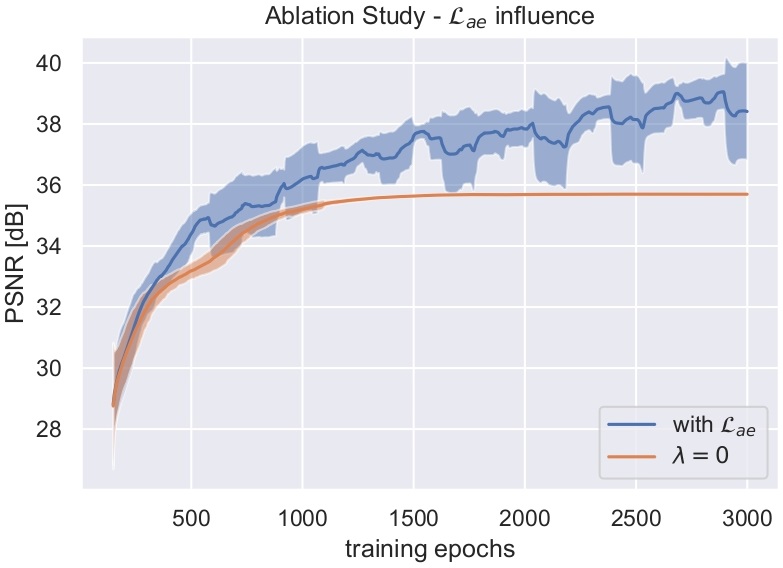}
    \caption{{Ablation study, influence of the $\bm{L}_{ae}$ loss function into the recovery performance. Color shadows correspond to confidence interval constructed from a moving standard deviation.}} \vspace{-1em}
    \label{fig:ae_loss}
\end{figure}

\section{Simulations on Synthetic Data}

\subsection{Experimental Setup}
The performance of the proposed method was evaluated on two hyperspectral image datasets (KAIST \cite{kaist}, and ARAD \cite{arad2020ntire}). Each dataset contained 31 spectral bands and was down-sampled to $512\times512$ spatial resolution. The datasets are composed of 30 and 460 images and were divided into 27/3 and 450/10 images, respectively, for training and testing \cite{monroy2021deep, arguello2021shift}. The simulated acquisition system was the Dual Disperser DD-CASSI. 

\break
   {Random-patch training is employed to avoid overfitting. This training scheme extracts 24 random spectral patches of $96\times96\times31$ size from each spectral image. Additionally, random coded apertures are independently drawn from a Bernoulli (1/3) distribution along training iteration to obtain an emulated 2D compressed projection for each spectral patch. The training configuration is composed of 3000 training epochs with 104 batch-size, $10^{-3}$ as learning rate with the Adam optimizer, running on  RTX 3090 GPU with 24 GDDR6 VRAM. Once the training procedure is finished,  the recovery of a whole spectral image is performed in a single forward step due to the translation invariance properties of the CAE.} {The quality of the spectral image estimation was evaluated using the peak signal-to-noise ratio (PSNR \cite{hore2010image}), structural similarity index measure (SSIM \cite{wang2004image}), and spectral angle mapper in radians unit (SAM \cite{kruse1993spectral})  metrics.}

\begin{figure}[!t]
    \centering
    \includegraphics[width=\linewidth]{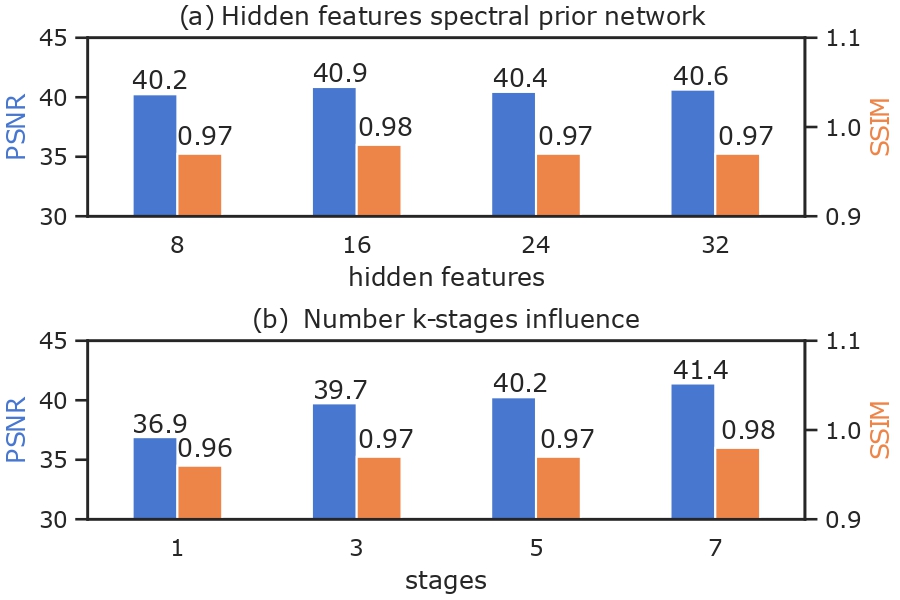} 
    \caption{Network configuration. (a) Influence of the number of hidden features for the spectral prior network, (b) Influence of amount of stages of the unrolled network.} 
    \label{fig:stagescomp}
\end{figure}

\begin{figure*}[!t]
    \centering
    \includegraphics[width=\linewidth]{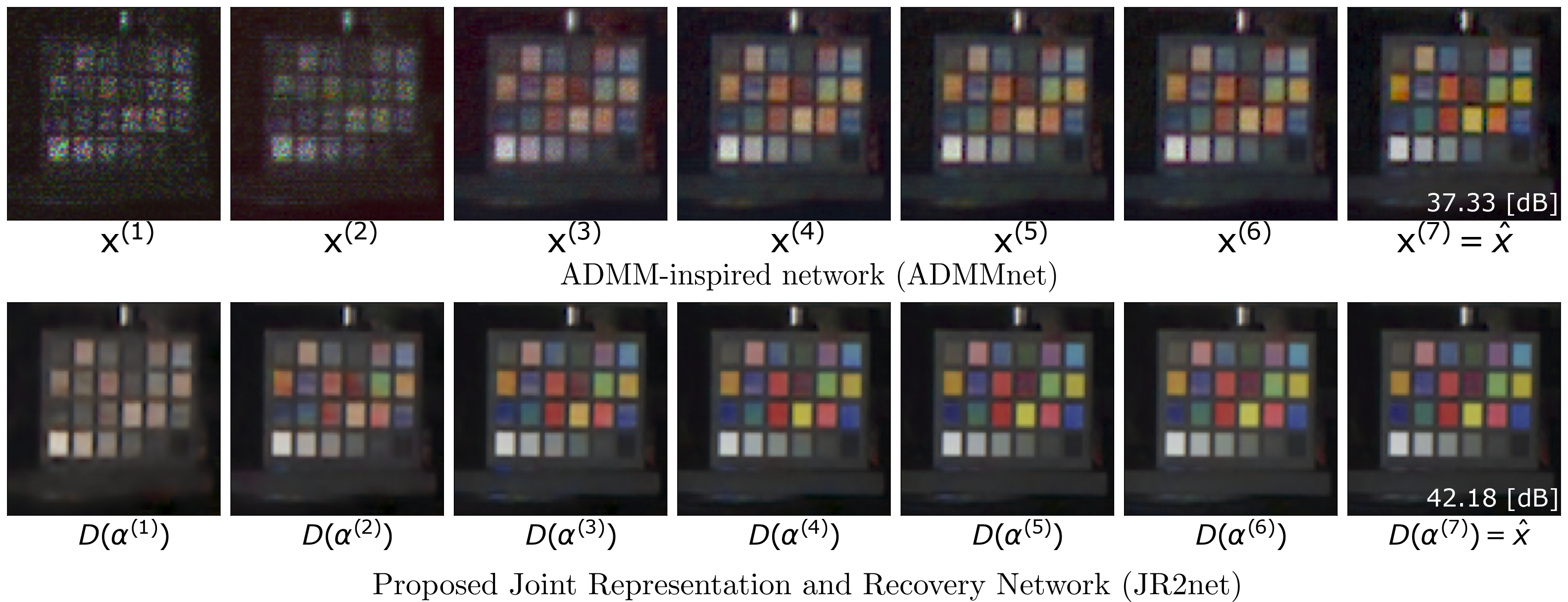} 
    \caption{Intermediate outputs along unrolled stages obtained by ADMMnet and proposed JR2net, with each column corresponding to the $k$-th stage estimation of the spectral image. The ADMMnet is the case when $\bm{D}(\cdot)$ and $\bm{G}(\cdot)$ are identity operators, namely $\bm{D}(\textbf{x}) = \bm{G}(\textbf{x}) = \textbf{Ix}$.}
    \label{fig:alongstages}    \vspace{-1em}
\end{figure*} 

\begin{figure*}[!t]
    \centering
    \includegraphics[width=\linewidth]{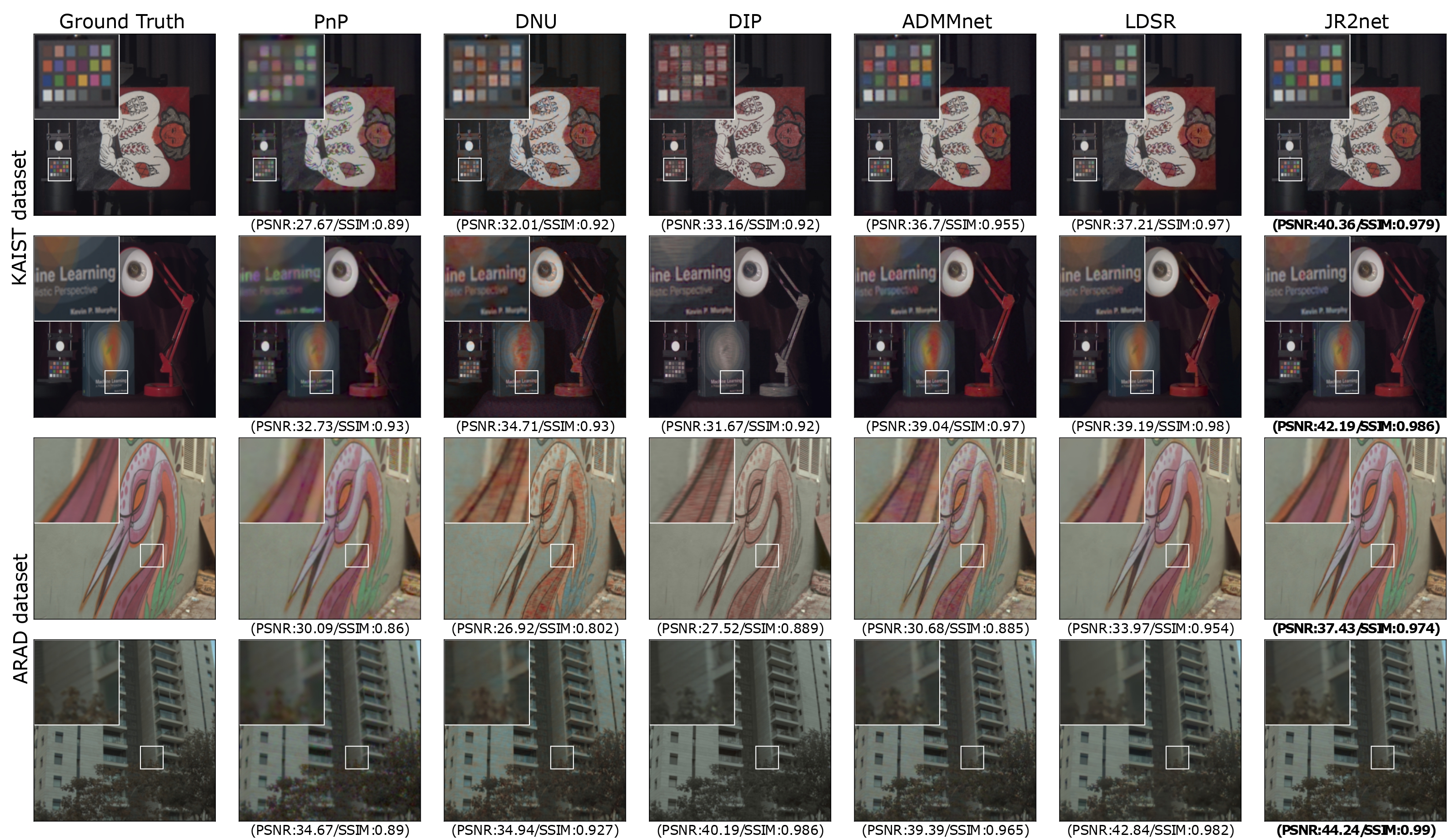}
    \caption{Recovery results: PnP \cite{yuan2020plug}, DNU \cite{dnu}, DIP \cite{dip_bacca}, ADMMnet \cite{sogabe2020admm}, LDSR \cite{monroy2021deep}, and the proposed method for two tested images on the  KAIST and ARAD datasets. The numbers in parenthesis show the averaged PSNRs and SSIMs scores.}  \label{fig:visualcom}  \vspace{-1em}
\end{figure*}

\subsection{Network Configuration}

The baseline configuration of the proposed network consists of a $\bm{D}(\cdot)$ and $\bm{G}(\cdot)$ of 5 hidden layers with an NLD representation of 8 features, as suggested in our previous work \cite{monroy2021deep}. The spectral prior network is a convolutional residual network following the architecture suggested by the authors in  \cite{kaist2}. To verify the effectiveness of introducing the $\bm{L}_{ae}$ into the training procedure of the NLD representation learning, an ablation study was conducted on the KAIST dataset. Experimental results show that introducing the $\bm{L}_{ae}$ loss function increases the recovery quality by around $3$ dB of PSNR, as shown in Fig.~(\ref{fig:ae_loss}). \textcolor{black}{Thus, it can be verified that regularizing the $\bm{G}(\cdot)$ network via the proposed loss function $\bm{L}_{ae}$ that simultaneously learns the derivatives of the decoder network and the non-linear transformation as an encoder network in CAE architectures, further increasing the global performance of the proposed network.}

Furthermore, the influence of the number of stages of the unrolled network and hidden features for the prior spectral network was evaluated on the PSNR and SSIM metrics. First, the proposed method was evaluated for a different number of hidden features $\{ 8, 16, 24, 32\}$ in the spectral prior network $\bm{H}(\cdot)$ with a fixed number of $5$ stages. As shown in Fig.~\ref{fig:stagescomp}(a), there is no significant increase in performance for hidden features greater than 16. Second, the proposed method was evaluated for different models of different numbers of stages with $ k = \{1, 3, 5, 7\}$ and a fixed number of $8$ hidden features. As can be seen in Fig.~\ref{fig:stagescomp}(b), as the number of stages increases, the performance of the recovery is increased, achieving the best performance for a value of $k = 7$. Therefore, based on simulation results, the network configuration with optimal cost and performance consists of $k = 7$ stages of the unrolled network and $16$ hidden features for the spectral prior network.

\begin{table}[H] \centering
\resizebox{\linewidth}{!}{%
\setlength\tabcolsep{0.1cm}
\begin{tabular}{@{}l|lll|lll|lll@{}}
\toprule
Metric  & \multicolumn{3}{c|}{PSNR}                                                 & \multicolumn{3}{c|}{SSIM}                                                 & \multicolumn{3}{c}{SAM}                                                  \\ \midrule
SNR     & \multicolumn{1}{c}{40} & \multicolumn{1}{c}{30} & \multicolumn{1}{c|}{25} & \multicolumn{1}{c}{40} & \multicolumn{1}{c}{30} & \multicolumn{1}{c|}{25} & \multicolumn{1}{c}{40} & \multicolumn{1}{c}{30} & \multicolumn{1}{c}{25} \\ \midrule
ADMMnet & 37.81                  & 37.34                  & 36.16                   & 0.960                  & 0.949                  & \textbf{0.925}          & 0.175                  & 0.222                  & 0.276                  \\
JR2net  & \textbf{40.29}         & \textbf{40.05}         & \textbf{37.13}          & \textbf{0.982}         & \textbf{0.965}         & 0.918                   & \textbf{0.115}         & \textbf{0.177}         & \textbf{0.248}         \\ \bottomrule
\end{tabular} } \vspace{-0.7em}
\caption{ Mean performance comparison for the proposed method and ADMMnet under different noise levels. \label{table:noise}  }
\end{table}

\subsection{Performance Evaluation}

We evaluated the influence of NLD representation learning. For this, the proposed method was compared for the case when the decoder and learned gradient network are the identity operator ($\bm{D}(\cdot)=\bm{G}(\cdot)=I$). In this case, the proposed method becomes the ADMMnet~\cite{sogabe2020admm}. Intermediate outputs of the spectral image estimations for the ADMMnet, as well as the proposed JR2net for the configuration of $k=7$ stages, are shown in Fig.~\ref{fig:alongstages}.  

\vspace{0.5em}

\begin{figure*}[b!]
    \centering
    \hspace*{-1cm}
    \fbox{\hspace{2em}  \includegraphics[width=\linewidth]{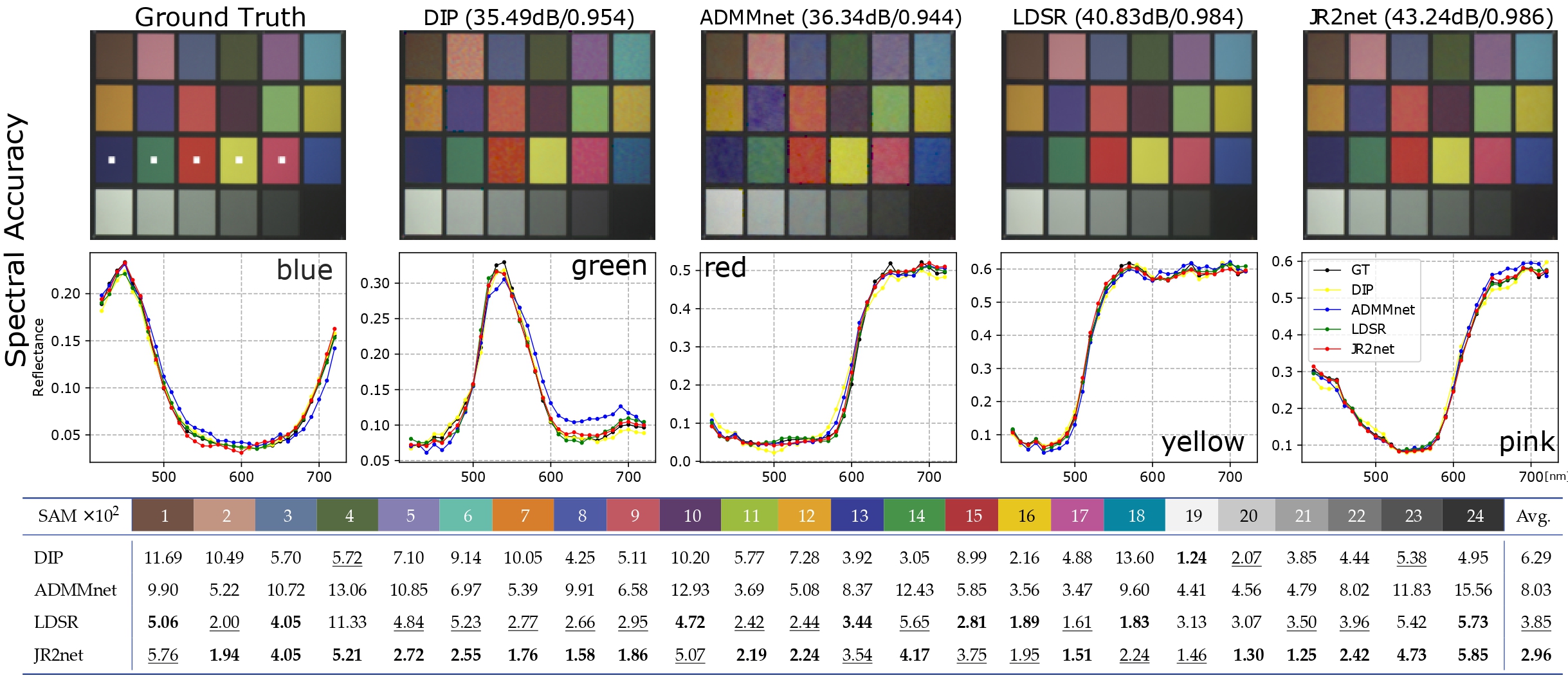} }
    \caption{Comparison of spectral accuracy of our spectral image recovery against three state-of-the-art methods: DIP, ADMMnet, and LDSR. Numbers in parentheses on top of the pictures show the average across spectral bands PSNR and SSIM scored between the recovery result and the ground truth. {The middle row shows pixel-wise recovered reflectances comparison for five color squares: blue, green, red, yellow, and pink. The bottom row shows the SAM score of the recovered spectra for the 24 color squares}.}
    \label{fig:spectral_acc} 
\end{figure*}

For ADMMnet results, the recovered image is gradually reduced in noise with the progression of the stages, which allows the $\bm{H}(\cdot)$ sub-network to be interpreted as a denoiser, which is in concordance of with the PnP methodology. Additionally, a good estimation can be observed from the first iteration, which shows an excellent estimation of the spatial information in the scene for the proposed method. The proposed model gradually improves spectral precision to achieve spectral image recovery. 

\vspace{0.5em}

Additionally, the ADMMnet method and the proposed method were simulated under different levels of additive Gaussian noise~$\{40, 30, 25\}$~dBs, with the simulation results shown in Table~\ref{table:noise}. The proposed method presents the best results in all the metrics for the proposed noise values, presenting optimal performance results for SNR values greater than 30 dB. In this sense, learning an NLD representation of spectral images allows one to obtain a better estimation in terms of spatial structure and estimation with less noise in the first stages of the algorithm, significantly improving the final performance.

\vspace{0.5em}

\begin{table}[!h]\centering
\resizebox{\linewidth}{!}{\begin{tabular}{l | ccc | ccc | c}
\toprule
Dataset & & KAIST &  &  &  ARAD & &  \multirow{2}{*}{GPU Time {[}s{]}}  \\\cline{1-7}
Metric  & PSNR & SSIM  &  SAM & PSNR & SSIM & SAM &   \\ \midrule
PnP         & 29.79             & 0.917             &  0.190            & 31.51             & 0.851             & 0.120         & 263.9\\
DNU         & 32.60             & 0.917             &  0.230            & 29.54             & 0.830             & 0.140         &  0.357 \\
DIP         & 32.79             & 0.920             &  0.220            & 35.78             & 0.958             & 0.080         & 441.5\\
ADMMnet     & 37.87             & 0.963             &  0.160            & 33.45             & 0.904             & 0.100         & \textbf{0.185}\\
LDSR        & \underline{38.84} & \underline{0.977} &  \underline{0.110}& \underline{38.23} & \underline{0.968} & \textbf{0.050}& 633.1\\
JR2net      & \textbf{41.41}    & \textbf{0.984}    &  \textbf{0.100}   & \textbf{39.65}    & \textbf{0.977}    & \textbf{0.050}& \underline{0.326}\\ \bottomrule
\end{tabular} }
\caption{ Performance comparison of the KAIST and ARAD datasets. The best performance is labeled in bold, and the second best performance is underlined.  } \label{table:performance} \vspace{-1em}
\end{table}

We compared the proposed method with state-of-the-art methods in compressive image recovery such as the Deep Non-Local Unrolling (DNU) \cite{dnu}, Plug and Play (PnP) \cite{yuan2020plug}, Deep Image Prior (DIP) \cite{dip_bacca}, ADMM-inspired network (ADMMnet)~\cite{sogabe2020admm}, and low-dimensional spectral reconstruction (LDSR) \cite{monroy2021deep}. \textcolor{black}{PnP, DIP, and LDSR are model-based optimization methods, and DNU and ADMMnet are deep-based optimization methods. DNU, PnP, and ADMMnet do not employ representation learning, while LDSR learns a non-linear representation detached from the recovery problem.}  All methods were implemented according to the hyperparameters suggested by the authors. Fig.~\ref{fig:visualcom} shows an RGB false color for three recovered spectral images on the KAIST and ARAD test datasets. Quantitative results are shown in Table~\ref{table:performance}; it can be seen that the proposed method outperforms the fastest method (ADMMnet) by 1.41~dB and 2.57~dB on KAIST and ARAD, respectively, and is around 2000 times faster than the second method with the best recovery accuracy (LDSR), which indicates a significant improvement in terms of accuracy and speed. The gain in recovery speed is because once the network is trained, the recovery of the spectral image is performed in a single forward step, requiring 0.326 seconds to recover each new measurement. Additionally, it is worth highlighting that the proposed method does not need to include a sparse representation constraint to achieve an optimal recovery quality, unlike the Wavelet basis used in LDSR.

\vspace{0.5em}

Finally, Fig.~\ref{fig:spectral_acc} compares the individual spectral accuracy of the state-of-the-art and the proposed method. The top row compares the ColorChecker spectral image from the KAIST dataset. The middle row shows the recovered spectral reflectances of five color squares in the chart: blue, green, red, yellow, and pink; our results are consistent with the spectral response of the ground truth. The table at the bottom shows the SAM score for each color square and the average. 

\begin{figure}[!b]
    \centering
    \includegraphics[width=\linewidth]{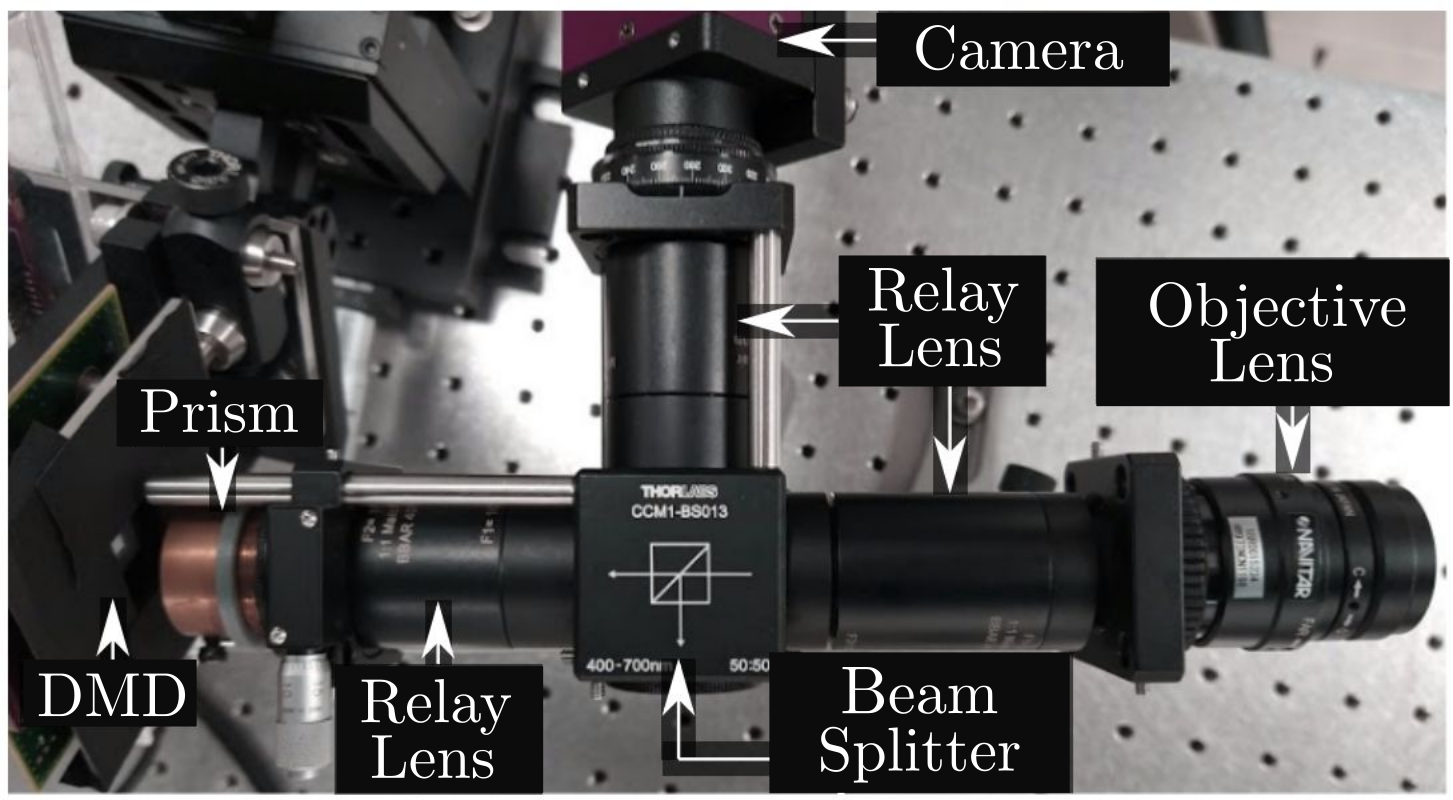}
    \caption{Testbed DD-CASSI imaging system implementation from authors in \cite{monsalve2021compressive}. {The sensing geometry performs the double dispersion of the incoming wavefront using a single dispersive element, a digital mirror device (DMD), and a beam-splitter (BS).}}
    \label{fig:realcassi}  \vspace{-1em}
\end{figure}

\subsection{Validation in Real Measurements}

\setcounter{figure}{10}
\begin{figure*}[b!]
    \centering 
    \hspace*{-1cm}  
    \fbox{ \hspace{2em} \includegraphics[width=\linewidth]{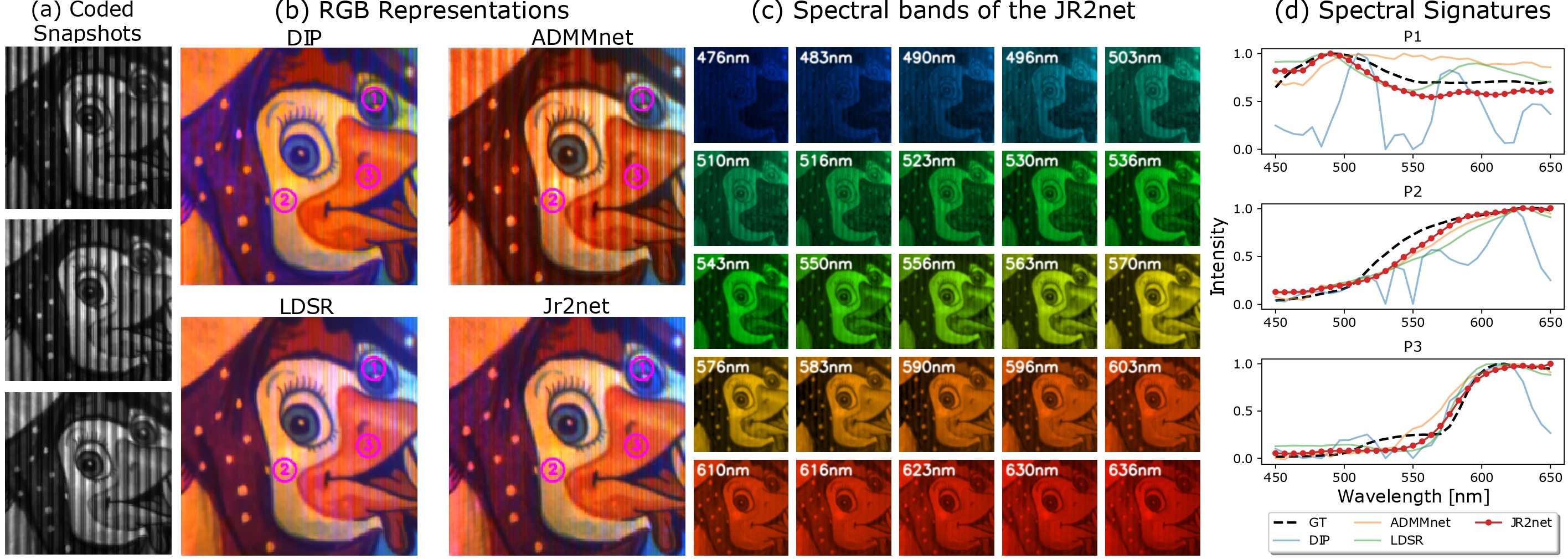} }\vspace{-0.5em}
    \caption{{(a) Coded Snapshots.} {(b)} RGB representation of the scene obtained with the evaluated recovery methods. {(c)} 25 Spectral bands of the recovered spectral image using the proposed method. {(d)} Normalized spectral signatures of three random spatial points are compared to a ground truth taken with a commercial spectrometer.} \vspace{-1.2em}
    \label{fig:real_visual} 
\end{figure*}

{We evaluated the proposed method with real measurements acquired from a testbed implementation of a spatial-spectral DD-CASSI imaging system provided by authors in \cite{monsalve2021compressive}.} The system uses a Navitar lens (12$mm$ FixedFocal Length, MVL12M23 - 12$mm$ EFL, $f$/$1.4$) as the objective lens, two matches achromatic doublet pair relay lens (Thorlabs MAP10100100-A, $f$1 $= 100.0mm$, $f$2 $=100.0mm$), a  double Amici prism coupled to a rotation mount (Thorlabs CRM1P, 30$mm$ cage rotation mount, 1"), a digital mirror device (DMD, Texas Instruments, D4120), a beam splitter (BS),  a third matched achromatic doublet pair relay lens (Thorlabs MAP105050-A, $f$1 = $50.0mm$, f$2$ = $50.0mm$), and a sensor (Stingray F-0980B, $4.65\mu m$ pixel size){, as shown in Fig.~\ref{fig:realcassi}. } {The experimental scene is illuminated by a reference 3900e DC regulated Lightsource with Light feedback.}


\break

The compressed projection and spectral image have a spatial resolution of $256 \times 256$ pixels with characterization to recover up to $C = 31$ spectral bands. The coded aperture designed in \cite{monsalve2021compressive} is generated from a random binary pattern with $18\%$ transmittance and acquires three snapshots. {As was detailed in \cite{monsalve2021compressive}, in the experimental setup, the coded apertures are placed in a commercial DMD, and a calibration process is carried out to obtain the calibrated sensing matrix. Because of the misalignment between the size of the DMD and the sensor and also due to the disperse effect of the prism, the resulted coded aperture is not binary, as shown in Figure~\ref{fig:codes}. Specifically, this calibration process consists of illuminating the system for different spectral bands in the range of 450nm to 650nm with 2nm spacing. Subsequently, spectral decimation is performed to obtain 31 desired spectral bands. Then, the calibrated sensing matrix obtained from the characterized coded aperture is used in the recovery algorithms.} Based on the performance evaluation summarized in Table~\ref{table:performance}, DIP, ADMMnet, LDSR, and the proposed method are selected for the real data validation. \vspace{-0.3em} 

\setcounter{figure}{9}
\begin{figure}[H]
    \centering
    \includegraphics[width=\linewidth]{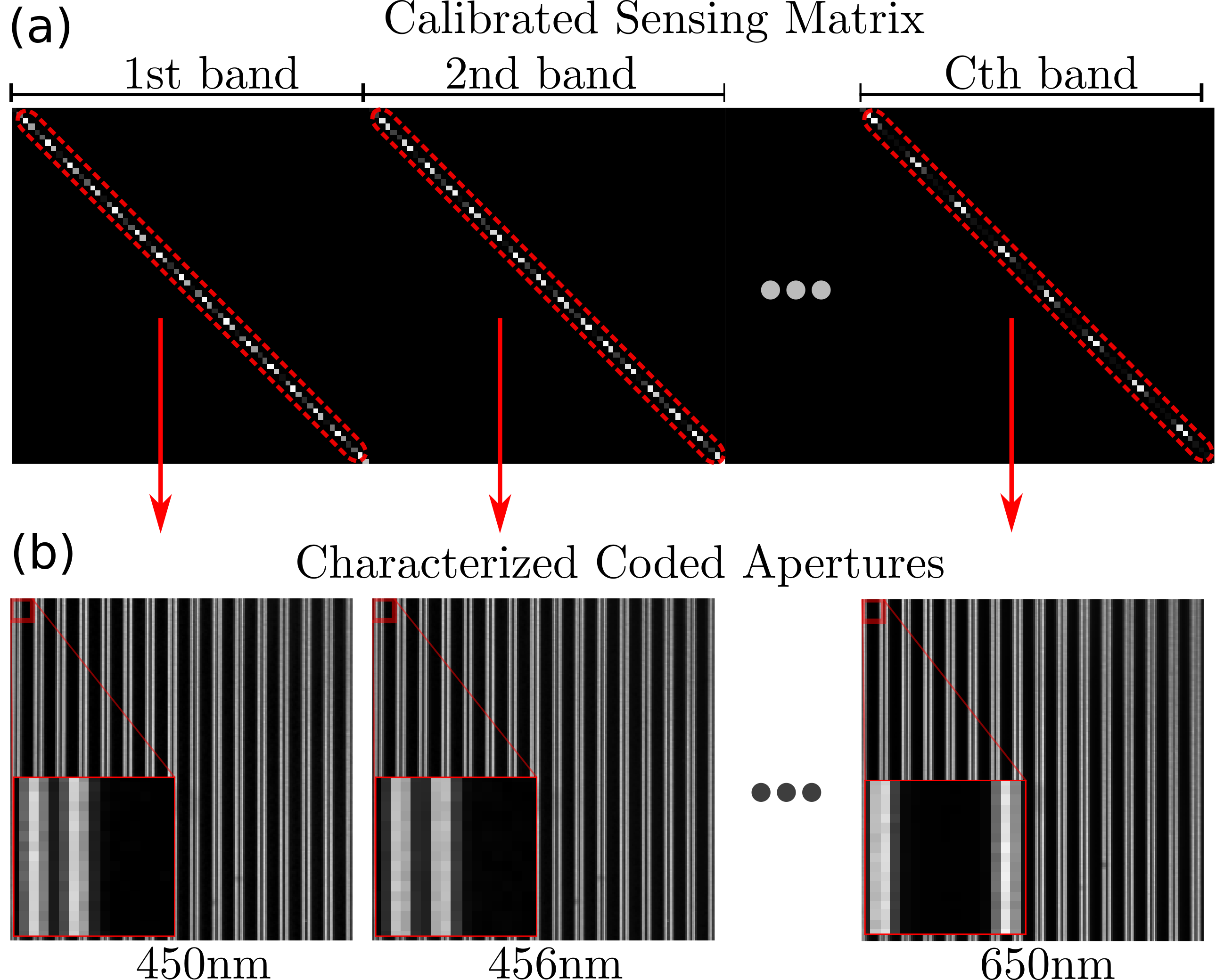} \vspace{-1.8em}
    \caption{{(a) Visual representation of the calibrated sensing matrix (b)~Characterized coded apertures for three spectral bands.} } \vspace{-0.3em}
    \label{fig:codes} \vspace{-1em}
\end{figure}

\break

{ Fig.~\ref{fig:real_visual}(a) presents the coded snapshots acquired from the DD-CASSI system.} Fig.~\ref{fig:real_visual}(b) shows a false-colored RGB of recovered spectral images obtained via the CIE RGB color matching functions \cite{wyman2013simple} for visualization. Fig.~\ref{fig:real_visual}(c) presents the spectral band representation of 25 spectral bands recovered for the proposed method. It can be seen that the proposed method obtains an optimal estimation of the spatial information. Furthermore, the spectral response of three particular spatial points in the scene, indicated as a magenta circle in the images is compared with a spectrometer (Ocean Optics Flame S-VIR-NIR-ES spectrometer), as shown in Fig.~\ref{fig:real_visual}(d). Our results show that the proposed method has a better estimation of the spatial and spectral information, and the spectral signatures are closer to the ones taken by the spectrometer. Notice that using an NLD representation in the unrolling network allows us to successfully estimate the response to the spectrum, unlike the ADMMnet method. The strategy of jointly learning the NLD representation with the recovery problem outperforms the pre-training approach. All computations were conducted using an NVIDIA Tesla K80 with 12GB of GDDR5 memory. The computation time of the JR2net was about 0.2 seconds. Code implementation can be found in~\cite{MonroyJR2net}.

\section{Conclusions}
This work proposes a joint non-linear representation and recovery network (JR2net) for the compressive spectral imaging problem. The JR2net contemplates the sensing systems and optimizes the learning of the non-linear low-dimensional representation for the recovery problem. The learned gradient network and the new autoencoder loss function reduce the computational complexity and increase recovery quality. This contribution relies on optimal estimations from the first stages of the unrolled network. Consequently, the proposed method outperforms state-of-the-art methods in terms of accuracy and speed, being around 2000 times faster than LDSR, the second method with the best recovery accuracy, and 2.57 dB above ADMMnet, the method with the best speed. The recovery results in real data that validates the performance of the proposed method for a DD-CASSI setup using a binary-coded aperture.

\begin{backmatter}

\bmsection{Funding} Universidad Industrial de Santander (VIE-project 2895, VIE-project 3720).

\bmsection{Acknowledgment}
The authors acknowledge the Vicerrectoría de Investigación y Extension of Universidad Industrial de Santander for supporting this work with projects  "Desarrollo de algoritmos de aprendizaje profundo basados en arquitecturas de extremo a extremo para la solución de problemas inversos" VIE-code 2895, and "Sistema óptico-computacional compresivo basado en aprendizaje profundo para la adquisición y fusión de imágenes con información espacio-espectral en el rango visible e infrarrojo cercano (400-1400 nm)" VIE-code 3720.

\bmsection{Disclosures}
The author declares no conflicts of interest.

\bmsection{Data availability} Data underlying the results presented in this paper are available in Ref. \cite{monsalve2021data}.

\bmsection{Supplemental document}
See Supplement 1 for supporting content. 
\end{backmatter}
\bibliography{references}

\bibliographyfullrefs{references}


\ifthenelse{\equal{\journalref}{aop}}{%
\section*{Author Biographies}
\begingroup
\setlength\intextsep{0pt}
\begin{minipage}[t][6.3cm][t]{1.0\textwidth} 
  \begin{wrapfigure}{L}{0.25\textwidth}
    \includegraphics[width=0.25\textwidth]{john_smith.eps}
  \end{wrapfigure}
  \noindent
  {\bfseries John Smith} received his BSc (Mathematics) in 2000 from The University of Maryland. His research interests include lasers and optics.
\end{minipage}
\begin{minipage}{1.0\textwidth}
  \begin{wrapfigure}{L}{0.25\textwidth}
    \includegraphics[width=0.25\textwidth]{alice_smith.eps}
  \end{wrapfigure}
  \noindent
  {\bfseries Alice Smith} also received her BSc (Mathematics) in 2000 from The University of Maryland. Her research interests also include lasers and optics.
\end{minipage}
\endgroup
}{}

\end{document}